\documentclass{aa}  

\usepackage[]{graphicx}
\usepackage{hyperref}
\usepackage[dvipsnames]{xcolor}
\usepackage{changes}
\usepackage{orcidlink}      %give your orcid
\usepackage{comment}
\usepackage{txfonts}
\usepackage{amsmath}
\bibpunct{(}{)}{;}{a}{}{,} % to follow the A&A style
\newcommand{\te}{\mbox {$T_{\mathrm{eff}}$}}
\newcommand{\sna}{SN~Ia}
\newcommand{\sne}{SNe~Ia}
\newcommand{\msun}{\mbox{$\mathrm{M_{\odot}}$}}
\newcommand{\rsun}{\mbox{$\mathrm{R_{\odot}}$}}
\newcommand{\lsun}{\mbox{$\mathrm{L_{\odot}}$}}

\newcommand{\rps}{\mbox {{\rm rad$\cdot$ s$^{-1}$}}}

\newcommand{\mdot}{\mbox{$\mathrm{\dot{M}}$~}}
\newcommand{\mdote}{\mbox{$\mathrm{\dot{M}}$}}
\newcommand{\msyrm}{\mbox{$\mathrm{M_{\odot} yr^{-1}}$}}
\newcommand{\isotope}[2]{${}^{#1}$#2}

\newcommand{\mwd}{\mbox {{\rm M$_{\rm WD}$}}}
\newcommand{\msd}{\mbox {{\rm M$_{\rm sdB}$}}}
\newcommand{\rwd}{\mbox {{\rm R$_{\rm WD}$}}}
\newcommand{\gcc}{\mbox {{\rm g$\cdot$cm$^{-3}$}}}

\newcommand{\ptf}{PTF~J2238}

\begin{document}

   \title{The expected evolution of the binary system PTF J2238+743015.1}

   \author{L. Piersanti\orcidlink{0000-0002-8758-244X}\inst{1,2}
        \and
          L.R. Yungelson \orcidlink{0000-0003-2252-430X} \inst{3}
        \and
          E. Bravo \orcidlink{0000-0003-0894-6450}\inst{4}
          }

   \institute{INAF-Osservatorio astronomico d'Abruzzo, via Mentore Maggini, snc, I-64100, Teramo, Italy\\
              \email{luciano.piersanti@inaf.it}
         \and
             INFN - Sezione di Perugia, via Pascoli, 123, I-06123, Perugia, Italy
         \and
             Institute of Astronomy of the Russian Academy of Sciences, 48 Pyatnitskaya Str, 119017 Moscow,
             Russia\\
             \email{lev.yungelson@gmail.com}
         \and
             Departamento de F\'{i}sica Te\'{o}rica y del Cosmos, Universidad de Granada, E-18071 Granada, Spain \\
             \email{eduardo.bravo@ugr.es}
             }

   \date{Received ; accepted }

% \abstract{}{}{}{}{}
% 5 {} token are mandatory
 
  \abstract
  % context heading (optional)
   {Binary systems harboring a low-mass CO WD and a He-rich donor are considered as possible progenitors of explosive events via He-detonation, producing low-luminosity thermonuclear Supernovae with a peculiar nucleosynthetic pattern. Recently, the binary system PTF~J223857.11+743015.1 has been suggested as one.} 
  % aims heading (mandatory)
   {We investigate the evolution of the PTF~J223857.11+743015.1 system, composed by a 0.75\,\msun\ CO WD  and a 0.390\,\msun\ subdwarf, capped by a thin H-rich layer, considering rotation of the WD component.}
  % methods heading (mandatory)
   {Using the FuNS code, we compute the evolution of two stars simultaneously, taking into account the possible evolution of the orbital parameters, as determined by mass transfer between components and by mass ejection from the system during Roche lobe overflow episodes. We consider that the WD gains angular momentum due to accretion and we follow the evolution of the angular velocity profile as determined by angular momentum transport via convection and rotation-induced instabilities.}
  % results heading (mandatory)
   {As the donor H-rich envelope is transferred, the WD experiences recurrent very strong H-flashes triggering Roche lobe overflow episodes during which the entire accreted matter is lost from the system. Due to mixing of chemicals by rotation-induced instabilities during the accretion phase, H-flashes occur inside the original WD. Hence, pulse-by pulse, the accretor mass is reduced down to 0.7453\,\msun. Afterwards, when He-rich matter is transferred, He-detonation does not occur in the rotating WD, which undergoes six very strong He-flashes and subsequent mass loss episodes. Also in this case, due to rotation-induced mixing of the accreted layers with the underlying core, the WD is eroded. Finally, when the mass transfer rate from the donor decreases, a massive He-buffer is piled-up onto the accretor which ends its life as a cooling WD.}
  % conclusions heading (optional), leave it empty if necessary
   {The binary system PTF J2238+743015.1 as well as all those binary systems having similar masses of the components and orbital parameters are not good candidates as thermonuclear explosions progenitors.} 


   \maketitle
%
%________________________________________________________________
\section{Introduction}\label{s:intro}
This paper continues a series of studies \citep{pier2014,piersanti2015,piersanti2019} devoted to investigate the consequences of He-rich matter accretion onto carbon-oxygen white dwarfs (CO WDs), which may occur in close interacting binaries harboring either a CO WD and a He WD or a CO WD and a nondegenerate He companion (a hot subdwarf - sdB).

As first recognized by \citet{taam1980a} and \citet{1980SSRv...27..563N}, within certain ranges of masses of CO WD and accretion rates of helium onto them in close binaries, accretion may produce the off-center detonation in the accumulated He layer,  which, in its own turn, possibly, may trigger detonation of carbon and result in a Type Ia supernova (\sna).  For a certain time this mechanism did not attract much attention, since the paradigm of close to Chandrasekhar mass \sna\ due to the merging of two CO WDs  dominated. It became popular anew, after \citet{livne90}, \citet{1991ApJ...370..272L} and other authors  demonstrated that He-detonation at the surface of a sub-Chandrasekhar WD may trigger converging shock waves which, in principle, are able to initiate detonation of underlying WD.
The latest studies \citep[e.g., ][]{2021ApJ...919..126B,Shen_2021,2023ApJ...951...28W} show that accretion of $\lesssim$0.03\,\msun\ of He onto a massive enough CO WD (\mwd$\ge$ 0.9\,\msun) is sufficient to initiate a double detonation and produce light-curves and spectra of most non-peculiar \sna. Detonations in more massive shells may result in peculiar  \sne. 
A comprehensive list of studies of ``double-detonation scenario'' (DD) and a discussion of their 
results may be found, e.g., in \citet{2023ApJ...951...28W}. In the latter paper it has been also demonstrated that in binary systems with a degenerate He-rich donor, the larger the entropy of the He WD, the lower the maximum attained mass transfer rate and, hence, the more likely the explosion of the CO WD. 

The overwhelming majority of works devoted to study the accretion of He-rich material onto CO WD neglects the effects of rotation on the thermal evolution of the accretor. However, \citet{yoon2004d} considered that the accreted matter could deposit angular momentum, ruling the WD spin-up, and investigated the effects of rotation-induced mixing and energy dissipation via viscous friction. Their results demonstrated that differential rotation results in heating of the accreted layers and a lower degree of degeneracy of the matter at the epoch of He-ignition, which occurs sooner as compared to models without rotation and accreting at the same mass transfer rate \mdote. This may prevent the development of a He-detonation, making He-burning to occur through a series of ``Helium Novae''. The outburst of V445~Puppis \citep{2003A&A...409.1007A} may be a manifestation of such an event. On the other hand, rather high masses of components in V445~Puppis imply a high \mdot not leading to a detonation even if the WD is not rotating. 
Additionally, \citet{yoon2004d} suggested that for low values of the accretion rate mass loss from the system may occur during outbursts. This could result in a detachment of the two components which, however, will be brought back into contact thanks to orbital angular momentum loss (AML) via gravitational wave radiation (GWR).  
Further, \citet{neunteufel2017} accounted for the possible generation of a magnetic field due to the spin-up produced by angular momentum deposition and included in their simulation the magnetic-induced instabilities. 

The studies of the influence of rotation on the outcomes of He-accretion pursued by Yoon, Langer and their collaborators found their climax in \citet{2019A&A...627A..14N}, who computed a grid of almost three hundred model binaries with masses of WDs between 0.54 and 1.1\,\msun\ and masses of nondegenerate He-donors from 0.4 to 1.0\,\msun, accounting for the effects of rotation and magnetic torques and with both components fully resolved. The grid was parameterized by initial orbital periods of the binaries, resulting in Roche lobe overflow (RLOF) by the model donors when He in their cores was exhausted to different extents. Such He-burning models of mass-transferring stars typically have \mdot$\approx 10^{-8} - 10^{-7}$\,\msyrm\ \citep{yung2008}. Evolution was traced up to the moment when the He-burning timescale became comparable to the WDs dynamical timescale. The main results of \citet{2019A&A...627A..14N} may be roughly summarized as follows: 1) it is necessary to accrete at least 0.1\,\msun\ of He to trigger a detonation; 2) He-detonation does not occur if $M_{\rm WD} \lesssim 0.8$\,\msun;
3) the masses of donors that provide mass-transfer rates low enough to initiate a detonation are inversely correlated with initial WD mass and span from 0.95\,\msun\ for \mwd=0.82\,\msun\ to 0.8\,\msun\ for \mwd=1.1\,\msun; 4) the range of initial orbital periods of binaries where detonation is possible is rather narrow, from $\lesssim 0.04$\,day for the less massive pairs to $\lesssim 0.06$\,day for the most massive ones.
In other systems either a He-deflagration occurs or He-burning in the donor ceases before a He-buffer massive enough to trigger He-ignition is accumulated onto the WD. In the latter case two CO WDs form.
 
The verification of theoretical modeling was hampered by an absence of observed close enough sdB+WD systems with well established parameters. However, a very close sdB+WD system was recently discovered and thoroughly studied by \citet{Kupfer2022}: PTF J223857.11+743015.1 (henceforth, \ptf). Another system, CD-30$^\circ$11223 (henceforth, CD-30) is known for more than a decade \citep{Vennes_2012}, but only recently the ambiguity in the determination of its parameters was resolved \citep[see][and references therein]{deshmukh2024}. For \ptf, it is expected that the subdwarf will overflow its Roche lobe in $\approx$6\, Myr. CD-30 is expected to become semidetached in $\approx$40\,Myr.

Following \citet{1990SvA....34...57T,2010ApJ...717.1006R}, \citet{Kupfer2022} assumed that the subdwarf was formed first as a result of the stable loss of most of the H-rich envelope of a (2 -- 5)\,\msun\ star with a non-degenerate He-core via Roche lobe overflow (RLOF) in the hydrogen-shell burning stage, while the CO WD formed via subsequent CE during the AGB phase.
Components of the binary are currently brought to contact due to AML via GWR. To satisfy $\log g$ and \te\ of the sdB, \citeauthor{Kupfer2022} assumed that the He-core of the donor has a $10^{-3}$\,\msun\ hydrogen envelope\footnote{Note, mass loss via RLOF from the progenitor of an $M_{\rm sdB} \simeq 0.4$\,\msun\  He-star in a close binary terminates, when the donor still has an $\approx$0.05\,\msun\ homogeneous H/He envelope and about same mass ``transition'' zone, where hydrogen abundance declines from initial one to zero. Mass-loss rate by hot subdwarfs \citep{2016A&A...593A.101K} is not sufficient to remove H/He envelope in 170 Myr, the estimated age of \ptf. One should postulate certain unknown process, which efficiently removes mass while pre-subdwarf crosses the HR diagram and changes chemical abundances profile.}. 

\citeauthor{Kupfer2022} used the  code  MESA \citep[][and subsequent papers]{2011ApJS..192....3P} to model the evolution of \ptf, approximating it by a binary system with \mwd=0.75\,\msun\ and \msd=0.41\,\msun. Possible rotation of the WD was not considered. Following \citet{woosley_kasen_dd11,2017ApJ...845...97B}, \citeauthor{Kupfer2022} predict that, after loss of the H/He envelope of the donor via accretion onto WD accompanied by a series of H-burning outbursts, accumulation of a $\Delta M=0.17$\,\msun\ He-layer on the WD surface should result in  a thermonuclear runaway ``that will likely destroy the white dwarf in a peculiar thermonuclear supernova''.

In the present study we compare the evolution of \ptf\, assuming non-rotating and rotating WD, by following in detail both  phases of H-rich and He-rich mass transfer. 
The paper is structured as follows. In  Section \ref{s:numerics} we describe the adopted input physics and the evolutionary code we use. In Section \ref{s:initial} the properties of the binary system after the formation of the compact configuration are described, while in Sections \ref{s:no-rot} and \ref{s:rot} the results of our computations for the non-rotating and rotating models of WD in \ptf\, are presented, respectively. In Section \ref{s:fate} we investigate the possible outcome of the \ptf\, evolution. In Section \ref{s:discuss} we discuss the uncertainties associated with the computation of the rotating model. Finally, in Section \ref{s:conclu} we draw our conclusions. 
%------------------------------- Table 1 
\begin{table*}
	\centering          
	\caption{Estimated properties of the binary system \ptf\, at the current epoch.}
	\label{bs_prop}
	\begin{tabular}{l r | l r | l r}     % 2 columns
		\hline\hline                 % inserts double horizontal lines
		\multicolumn{2}{c|}{\it Orbital Parameters} & \multicolumn{2}{c|}{\it Accretor} & \multicolumn{2}{c}{\it Donor}\\
		\hline
		P [min.]                       & 76.4     & \mwd\, [\msun]               & 0.75   & \msd\, [\msun]            & 0.390   \\
		A [\rsun]                     & 0.6208   & \rwd\, [$10^{-2}$\rsun]      & 1.082  & ${\rm R_{sdB}}$ [\rsun]            & 0.1851  \\
		$q={\rm \frac{M_{sdB}}{M_{WD}}}$  & 0.52     & ${\rm T_{eff}}$ [K]                & 26281  & ${\rm T_{eff}}$ [K]                & 23650   \\
		\multicolumn{2}{c|}{\it }                 & L [\lsun]                  & 0.050  & L [\lsun]                  & 9.652   \\
		\multicolumn{2}{c|}{\it }                 & \multicolumn{2}{c|}{\it }              & $\log(g)$                    & 5.48    \\
		\hline
	\end{tabular}
\end{table*}

\section{Input Physics and Numerics}\label{s:numerics}
All our models have been computed by an updated version of the hydrostatic 1D Lagrangian code FuNS \citep{straniero2006,cristallo2009,piersanti2013}. With respect to the previous version, we added the possibility of following simultaneously the evolution of both components in a binary system to evaluate self-consistently the mass transfer rate and the corresponding evolution of the orbital parameters. In this regard we assume that, if during a mass transfer episode the accretor fills its Roche lobe, the matter is lost from the system with the specific angular momentum of the accretor  \citep{piersanti2015,piersanti2019}. Here, we consider models having initial solar metallicity, adopting the solar mixture as derived in \citet{piersanti2007} for the compilation of solar abundances provided by \citet{grevesse1998}. We adopt the equation of state provided by \citet{stra1988} and successive upgrades \citep{prada2002} for temperature larger than $3\times 10^6$ K, while for lower temperature the tables of thermodynamic variables EOS\_2005 are used \citep[\url{https://opalopacity.llnl.gov/EOS_2005/}, ][]{rogers1996}. 
Radiative opacities for temperature lower than $5\times 10^8$ K have been obtained via the web facility provided by the OPAL group \citep[\url{http://opalopacity.llnl.gov/new.html}, see ][]{igles1996}, while at larger temperature  we make use of the LANL OPLIB tables \citep[\url{http://aphysics2.lanl.gov/opacity/lanl}, see ][]{colgan2018}. Conductive opacities are included according to the prescription in \citet[][\url{http://www.ioffe.ru/astro/conduct/index.html}]{cassisi2021}.
As in \citet{pier2014}, the adopted nuclear network includes elements from H to Fe, linked by $\alpha$-, $p$- and $n$-capture reactions as well as $\beta^\pm$ decays. To take into account the energy contribution due to $n$-captures on isotopes heavier than \isotope{56}{Fe}, we introduced a  fictitious neutron sink following the prescription by \citet{jorissen1989}. At variance with our previous works on CO WDs accreting matter from He-rich companions, we include the NCO chain by adopting the weak rates for $e$-capture on \isotope{14}{N} and $\beta$-decay of \isotope{14}{C} provided by \citet{PCpinedo}, while the reaction rate for \isotope{14}{C}(${\alpha}$,${\gamma}$)\isotope{18}{O} process is derived from \citet{johnson2009}.

In the computation of rotating models, the effects of rotation are accounted for as in \citet{piersanti2013} (see their Section 2), by assuming that the transport of angular momentum could be represented as a pure diffusive process. Responsible for angular momentum transport we consider convection (which is assumed to enforce rigid rotation in convectively unstable regions), as well as rotation-induced circulations, such as Eddington-Sweet (ES), Goldreich-Schubert-Fricke (GSF), Solberg-H\"oilland (SH) and shear (both dynamical and secular - DS and SS, respectively) instabilities. Hence, the total diffusion coefficient for the transport of angular momentum $D_j$ is given by 
\begin{equation}
	D_j=D_{conv}+D_{ES}+D_{GSF}+D_{SH}+D_{SS}+D_{DS}, 
	\label{diffj}
\end{equation}
where $D_{conv}, D_{ES}, D_{GSF}, D_{SH}, D_{DS}$\ are computed as described in detail in \citet{piersanti2013} \citep[see also ][]{endal78,heger2000}, while $D_{SS}$ is computed according to \citet{zahn1992}, as implemented by \citet[][- see their Eq. (13)]{yoon2004a}. 
We evaluate the electron viscosity, necessary to define the onset of secular shear instability, according to \citet[][ and references therein]{itoh1987}. We treat the presence of $\mu$-barriers by defining an equivalent $\mu$-current opposing to the rotation-induced circulation. Following \citet{heger2000}, the $\mu$-gradient $\nabla\mu$ in all the quantities defining the diffusion coefficients above is replaced by $f_\mu\nabla\mu$ where $f_\mu$ is a free parameter in the range [0,1] describing the sensitivity of rotation-induced instabilities to chemical gradients. Modifications of the local chemical composition due to both convection and rotation instabilities are described by adopting a diffusive scheme. The corresponding total diffusion coefficient $D_M$ becomes now:
\begin{equation}
	D_M=D_{conv}+f_c\cdot(D_{ES}+D_{GSF}+D_{SH}+D_{SS}+D_{DS}). 
	\label{diffc}
\end{equation}
where $f_c$ is a parameter in the range [0,1] describing the mixing efficiency of rotation-induced instabilities. 
At variance with \citet{piersanti2013}, we set $f_c$=0.04, as derived by \cite{pinsonneault1989} to reproduce the surface abundance of lithium in the Sun, and $f_\mu$=0.05, as derived by \citet{heger2000} to reproduce the nitrogen surface abundance in massive stars. 

Following \citet{kippen1978}, we include in the energy conservation equation the dissipation of rotational energy via the friction term $\varepsilon_{diss}$. Following \citet{yoon2004e} and  \citet[][ see their Appendix B.6]{paxton2013}, the  finite difference equations describing the transport of angular momentum are solved implicitly by using multiple sub-timesteps $\delta t_i$ smaller than the timestep $\Delta t$ used to integrate the other stellar equations. The exact values of $\delta t_i$ are fixed so that they appear smaller than the minimum local diffusion timescale related to the angular momentum transport. Such a procedure is adopted to improve the accuracy in evaluating the angular velocity profile.

In the present work we assume that H or He are ignited when the energy released per unit time by CNO-cycle or, respectively, 3$\alpha$-reactions first becomes higher than the surface luminosity by a factor 100. Consistently, we define the H- or He-ignition point as the mass coordinate where H- or He-burning has a maximum at the ignition epoch.

\section{The Initial System}\label{s:initial}
Basing on the analysis and results of \citet{Kupfer2022}, we model the current epoch \ptf\, as a binary harboring a CO WD of 0.75\,\msun\, and an sdB companion of 0.390\,\msun\ with an orbital period $P=76.4$ min. After \citeauthor{Kupfer2022}, we assume also that the compact binary formed 172 Myr ago with an orbital period of $P_0=122.7$ min. and has a He-burning core of 0.3879\,\msun\ and an H-rich outer layer of $2.1\times 10^{-3}$\,\msun.

The initial CO WD model for our computations has been constructed by accreting He-rich matter onto the ``heated'' model M070 from \citet{pier2014} at \mdot$=3\times 10^{-7}$\,\msyrm, to simulate the evolution in the AGB phase. With this choice of parameters we find that at the current epoch the \ptf\, system has the properties presented in Table~\ref{bs_prop}, in agreement with those derived observationally by \citet{Kupfer2022}. By computing the expected further evolution of this system, we find that the sdB component will fill its  Roche lobe in 13.08 Myr. 

\section{Non-Rotating Model}\label{s:no-rot} 

\begin{figure}[t]
	\centering
	\includegraphics[width=\columnwidth]{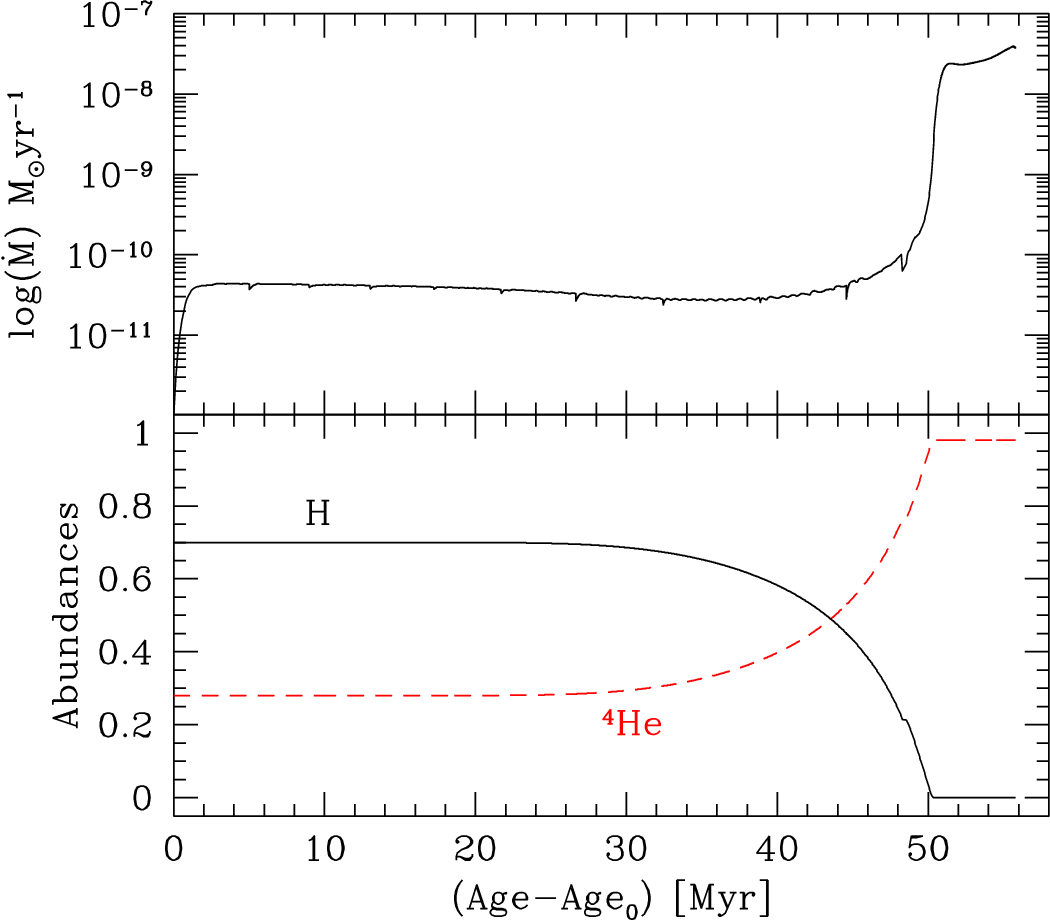}
	\caption{Time evolution of the mass transfer rate (upper panel) and of the H- and He-abundances in the transferred matter (lower panel) for the non-rotating model. The origin of the \textit{x-axis} has been fixed at the epoch when the sdB component filled its Roche lobe (Age$_{\rm 0}$=185.08\,Myr).
	}
	\label{fig1}
\end{figure}

\begin{figure}[h]
	\centering
	\includegraphics[width=\columnwidth]{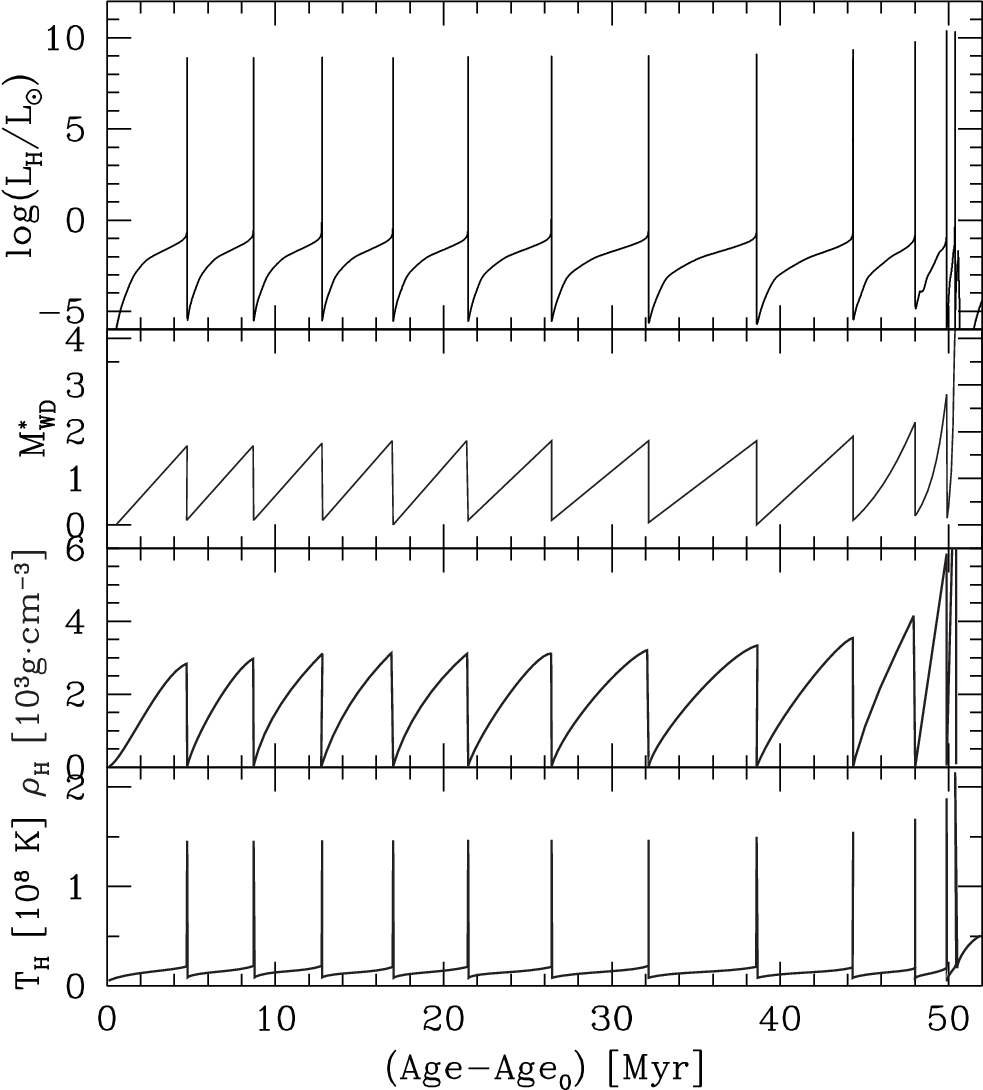}
	\caption{Evolution of selected physical properties of the non-rotating CO WD during the accretion phase of H-rich material. From top to bottom we report: the luminosity of the H-burning shell (${\rm {L_H}}$ in solar units); the variation in the CO WD total mass defined as ${\rm M_{WD}}^*=10^4\times\left({\frac{M_{WD}}{M_\odot}}-0.75\right)$; the density ($\rho_H$ in $10^3$ \gcc) and the temperature ($T_{\rm H}$ in $10^8$\,K) at the base of the H-burning shell.  
	}
	\label{fig2}
\end{figure}

First, we compute the evolution of the \ptf\, system neglecting the effects of rotation. This model will be used as a reference for the analysis of the rotating model (see the next Section). 
In the upper panel of Fig.~\ref{fig1} we present the history of mass transfer from the sdB  component onto the CO WD, while in the lower one we report the hydrogen (solid black line) and helium (dashed red line) abundances (by mass fraction) as tracers of the chemical composition of the accreted matter. 
In Fig.~\ref{fig2} several physical properties of the accreting WD are reported for the phase of H-rich matter accretion. During this phase, the CO WD experiences eleven H-flashes followed by a Roche lobe overflow. The mass transfer rate is very low, so H-flashes are very strong and all the matter accreted on the WD is lost from the system\footnote{Though outbursts are strong, we expect that they do not have Novae scale, since to get a Nova it is necessary to largely increase the CNO abundance at the base of the H-burning shell. The only possible way is mixing with the underlying WD, but the CO core is capped by a He-buffer whose outermost zone has an \isotope{14}{N} abundance equal to that of the initial CNO abundance, and this does not suffice to trigger a Nova.}.
%------------------ Figure 1  -----------------------------------------------

\begin{figure}
	\centering
	\includegraphics[width=\columnwidth]{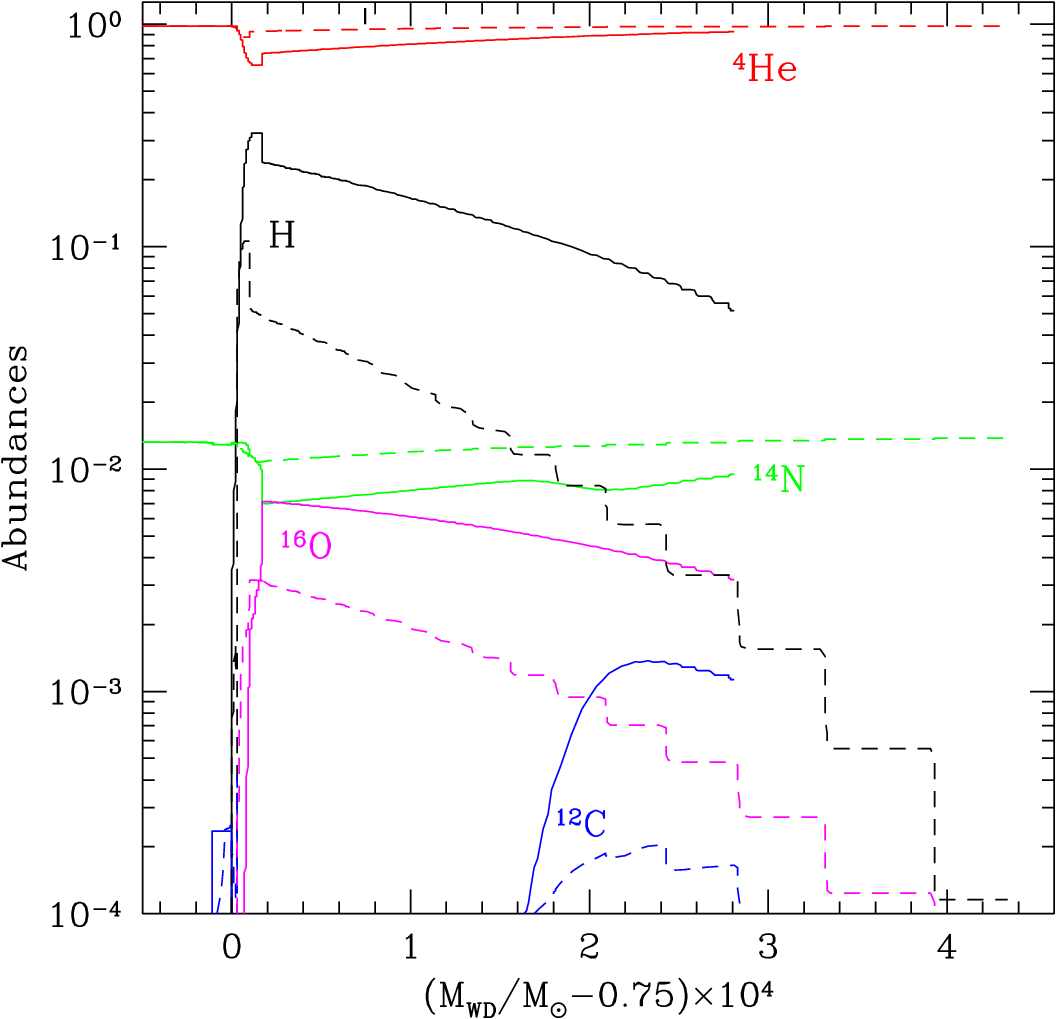}
	\caption{Comparison of the abundances of selected isotopes involved in the CNO cycle for  the non-rotating model at the onset of the H-flash in the eleventh (solid line) and twelfth (dashed line) cycles, respectively.}
	\label{fig3}
\end{figure}

A further inspection of Fig.~\ref{fig2} suggests that for the first eight H-flashes the amount of mass to be accreted to trigger the flash episode slightly increases pulse-by-pulse, because the mass transfer rate decreases (see the upper panel in Fig.~\ref{fig1}). 
After that, \mdot starts increasing, since H-abundance in the accreted matter declines. This results in the decrease of the interpulse periods.
%------------------ Figure 2  -----------------------------------------------
The last,  12$^{th}$ H-flash at $\simeq$50.4 Myr after the onset of the mass transfer is peculiar in many regards. Compared to the previous flash, the average mass transfer rate is about two orders of magnitude larger ($2\times 10^{-9}\ \mathrm{vs.}\ 2\times 10^{-11}$\,\msyrm), while the amount of mass that is accreted and triggers the flash also almost doubles ($4.35\times 10^{-4}\ \mathrm{vs.}\ 2.64\times 10^{-4}$\,\msun), so that the density at the  base of the H-shell at the onset of the thermonuclear runaway is also larger ($8.92\times 10^3\ \mathrm{vs.}\ 4.94\times 10^{3}$\gcc). Notwithstanding, the maximum luminosity (\textit{i.e.} the maximum value of the total nuclear energy released per unit time via H-burning) during the two flashes is practically the same, $L_H\simeq 10^{10}$\,\lsun, and, most important, the mass lost during the Roche lobe triggered by the 12$^{th}$ flash is only $2.6\times 10^{-5}$\,\msun\, (not visible in the plot). 
All these facts are due to the circumstance that during the twelfth accretion cycle of H-rich matter the H-abundance is very low, as shown in Fig.~\ref{fig3}, where we report the chemical stratification of the accreted layer at the onset of the eleventh and twelfth H-flashes. This implies that during the very last H-flash the amount of energy per unit mass injected in the accreted layers is small and, hence, the external layers of the accreting WD do not need to inflate very much to restore an equilibrium configuration.
%------------------ Figure 3  -----------------------------------------------

After the H-rich layer of the sdB component has been completely lost, He-rich material is accreted onto the WD (see Fig.~\ref{fig1}). We find that when the total mass of the WD becomes \mwd=0.888\,\msun, i.e. after a He-buffer as massive as 0.138\,\msun\, has been accumulated, a He-flash will be ignited at the mass coordinate 0.8\,\msun, where the density at the ignition epoch is $8\times 10^5$\gcc. As the FuNS code is a hydrostatic one, we check the possibility that this model could produce a successful explosion by using the 1D SN Ia explosion code Thermonuclear SUpernova Hydrodynamics and Nucleosynthesis \citep[TSUHN - ][]{bravo2019}. We find that the entire WD is destroyed, the kinetic energy of the ejecta being $1.07$ foe. As a consequence of the explosion, 0.295\,\msun\ of \isotope{56}{Ni} are produced, 0.260\,\msun\ from the CO core and the remaining 0.035\,\msun\ from the He-rich mantle. Among intermediate mass elements, we find that the produced total amount of sulfur and calcium are 0.122\,\msun\ and 0.04\,\msun, respectively, while the mass of ejected manganese is 0.00187\,\msun. 
The chemical composition of the ejecta is peculiar, as the abundances relative to \isotope{56}{Fe} of \isotope{78}{Kr}, \isotope{80}{Kr}, \isotope{44}{Ca}, \isotope{48}{Ti}, \isotope{74}{Se}, and \isotope{51}{V} are more than 10 times larger than the solar values.

Due to the explosion, it is expected that the remnant of the donor in \ptf, having a final mass of \msd=0.2496\,\msun, is kicked off with an escape velocity equal to its orbital velocity ${\rm v}_{sdB}=867$ $\mathrm{km\cdot s^{-1}}$.

The amount of accreted He-rich matter and the WD mass at detonation are commensurate with the values obtained for the non-rotating model of \ptf\, by \citet{Kupfer2022} -- $\Delta M_{\rm He}=0.17$\,\msun, \mwd=0.92\,\msun, respectively. As well, it reasonably agrees with data of \citet{2016A&A...589A..43N}, who found that for (0.8 --1.0)\,\msun\ WD, on average, 0.163\,\msun\ of He should be accumulated for detonation.

We remark that the occurrence of He-detonation is related to the mass transfer history  in the stage of He-rich matter accretion, which, in turn, depends on the adopted binary parameters for \ptf\, at the current epoch. In particular, for a  different set of the donor and/or accretor masses, as well as of the separation, \mdot may increase with time more rapidly than the one reported in Fig.~\ref{fig1}. Helium would then be ignited via very strong non-dynamical flash. In this case the binary will survive and the accreting WD will experience several He-flashes, during which all the accreted matter will be ejected from the system via RLOF. In this case, the fate of the WD should be a CO core surrounded by an extended He buffer \citep[see the discussion in][]{piersanti2015,brooks2015}.

\section{Rotating Model}\label{s:rot}

According to the observations of a sample of DA WDs \citep[][]{berger2005}, the rotation velocity of compact degenerate stellar remnants is very low ($\omega\le 10^{-2}$\,\rps), clearly suggesting that during the post-main-sequence evolution angular momentum has been subtracted from the radiative core. 
This conclusion is confirmed by the rotational curve derived from the data obtained by the \textit{Kepler} mission for a sample of Red Giant Branch stars \citep{mosser2012}, showing that the ratio of the angular velocity in the radiative core and the convective envelope is in the range 5-10 \cite[see, for instance, the discussion in ][]{eggenberger2012,marques2013}. Hence, we can safely assume that both compact objects in \ptf\, are initially practically at rest. 
Moreover, by assuming that, after the formation of the compact binary system, the synchronization of the orbits occurs on a very short timescale and remembering that the orbital period of the system under study in the course of further evolution will be in the range 10-70 min., it comes out that the angular velocity of both components before RLOF by the subdwarf will remain always lower that 0.01 \rps, so that the correction to the local gravity, proportional to $(\omega_{WD}/\omega_c)^2$, is lower than 1\% \cite[see also the discussion in Sect. 5 of][]{Bauer2021}. According to these considerations, we can safely assume that the sdB evolution is not affected by rotation. On the other hand, it is expected that during the mass transfer process, the accreted matter supplies angular momentum to the CO WD, which will become very soon a fast rotator, at least in the external zone. So in our calculation of the evolution of  \ptf\, we include the effects of rotation in the computation of the accretor only. 

For the mass ratio of donor and accretor in the model under study, mass transfer is stable \citep{2004MNRAS.350..113M} and the material transferred from the donor forms a thin Keplerian accretion disk around the accretor. Matter deposited onto the WD surface carries an amount of angular momentum almost equal to the Keplerian value at the equator. This implies that very soon after the onset of mass transfer the CO WD surface attains the critical rotation rate and, hence, in principle, no more matter can be added. \citet{pacz1991} and \citet{popham1991} demonstrated that under these conditions mass continues to be accreted onto the WD, which mantains its angular velocity close to the break-up value and transfers back to the disk the angular momentum excess. Hence, following \citet{yoon2004a}, we assume that the accreted matter does not deposit angular momentum if the rotation velocity on the WD surface is equal to its  Keplerian value:
   \begin{equation}
      j_{\rm acc} = \left\{\begin{array}{ll}
                           f_j \cdot j_{K} & \textrm{if $v_{\rm WD} < v_K$ } \\
                                         0 & \textrm{if $v_{\rm WD} = v_K$,} 
                           \end{array} \right.
    \label{eqjm}
   \end{equation}
where $v_{WD}$ is the surface equatorial velocity of the accreting WD, $v_K=\omega_K\cdot R_{\rm WD}=\sqrt{GM_{\rm WD}/R_{\rm WD}}$ is the Keplerian velocity at the WD equator, $j_{\rm acc}$ is the specific angular momentum of the accreted matter and $j_K=v_K R_{WD}=\sqrt{GM_{WD}R_{WD}}$ is the specific angular momentum at the WD equator. In our computation we fix the factor $f_j$ in Eq. (\ref{eqjm}) to 1. 
%------------------ Figure 4  -----------------------------------------------
\begin{figure}
	\centering
	\includegraphics[width=\columnwidth]{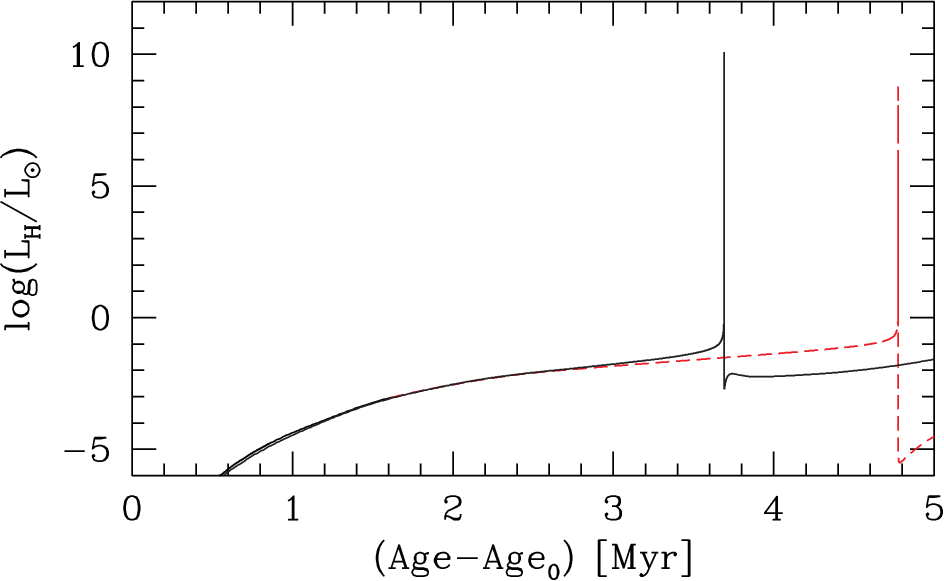}
	\caption{Time evolution of the H-burning shell luminosity during the first H-flash for the non-rotating and rotating models (red-dashed and black-solid lines, respectively).  
	}
	\label{fig4}
\end{figure}

\subsection{Evolution during the H-rich matter accretion phase}

In Fig.~\ref{fig4} we report for the rotating model (solid black line) the time evolution of the H-burning luminosity during the first H-rich matter accretion episode; for comparison we report also the same quantity for the non-rotating model discussed in \S \ref{s:no-rot} (red dashed line). As the mass transfer history is the same for the two models, it comes out that in the rotating model the first H-flash is ignited sooner, \textit{i.e.} when a smaller amount of mass has been accreted ($\Delta M_{acc}=1.27\times 10^{-4}$\,\msun\, and $1.74\times 10^{-4}$\,\msun\, for the rotating and non-rotating model, respectively). This is somewhat expected and is related to the heating induced by the rotational energy dissipation due to the transport of angular momentum via rotation-induced instabilities \citep[see the discussion in ][]{yoon2004d}. Figure~\ref{fig4} also discloses that the H-flash is stronger in the rotating model (the maximum $L_H$ value is larger than the one in the non-rotating model). Further analysis reveals that the density at the H-ignition point in the rotating model is $\rho^r_H=3.59\times 10^3$\gcc, a factor almost 2 larger than in non-rotating model ($\rho^{nr}_H=1.98\times 10^3$\gcc).
%------------------ Figure 5  -----------------------------------------------
\begin{figure}
	\centering
	\includegraphics[width=\columnwidth]{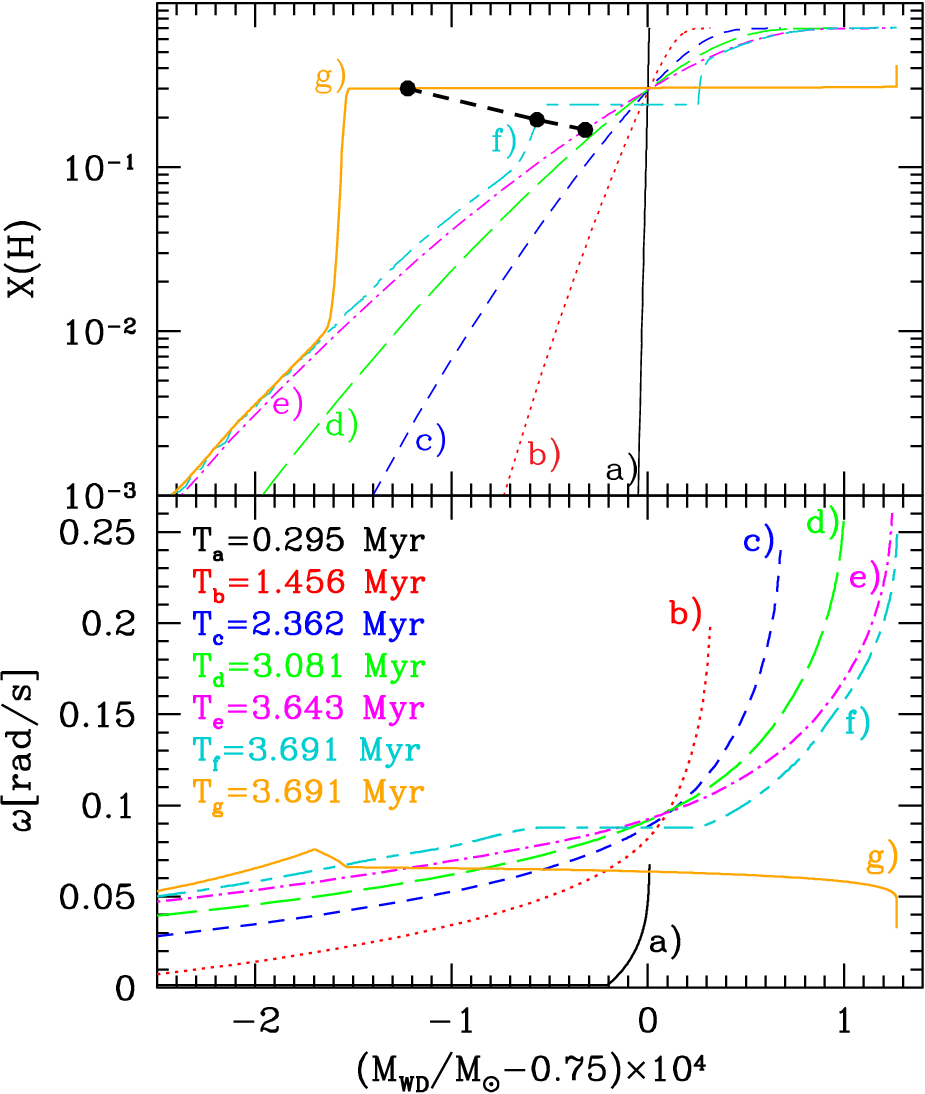}
	\caption{Profiles of hydrogen abundance X(H) (upper panel) and angular velocity (lower panel) at various epochs during the first mass transfer episode in the rotating model. The time after the onset of mass transfer corresponding to each line is reported in the plot. The \textit{x-axis} is defined in such a way that the original WD surface (M=0.75\,\msun) coincides with the origin. Lines \textit{a} to \textit{e} correspond to the evolution before the onset of H-flash; line \textit{f} corresponds to the epoch when flash-driven convection sets in; line \textit{g} corresponds to the epoch when flash-driven convection attains its maximum extension. Dots connected by heavy dashed line mark the position of the H-burning shell. 
	}
	\label{fig5}
\end{figure}

In the non-rotating model H-flash occurs at the surface of the original CO WD. Since rotation makes less dense the external layers of the accreting WD, in the rotating model the first H-flash occurs well below the surface of the original WD. Such an occurrence is due to the mixing induced by rotation-driven instability (mainly shear) of the accreted layer with the most external zone of the original WD. This is clearly illustrated in Fig.~\ref{fig5} where we report, for various epochs during the first mass transfer episode of H-rich material, the H-abundance X(H) as a tracer of the accreted matter (upper panel) and the angular velocity $\omega$ (lower panel) as a function of the WD mass.
As it can be noticed, due to rotation-induced mixing, the H-rich zone, \textit{i.e.} the zone of the WD where X(H) >0.1, during the H-flash is a factor 2.2 more massive than the accreted one (orange solid line, labeled \textit{g}).
A further inspection of Fig.~\ref{fig5} discloses that at the epoch of maximum extension of flash-driven convective shell, the zone below the H-burning shell has a flat 
distribution of chemicals (see line \textit{g}), a peculiarity as compared to flashes occurring in non-rotating models . In fact, after the onset of convection, shear instability transports inward angular momentum and mixes efficiently the zone above the H-burning shell with the underlying ones. As a consequence, also these inner layers become unstable for convection. As the flash proceeds, the local increase in temperature due to the H-flash splits the convective shell so that matter is no more carried efficiently through the burning shell.
To investigate further this issue, we recompute this first H-flash episode by setting in Eq.~\ref{eqjm} $f_j=$0.6, \textit{i.e.} by assuming that the accreted matter has a specific angular momentum equal to 60\% of the equatorial Keplerian value. We find that, with respect to the rotating model described above, at the H-ignition (which occurs at M=0.7499\,\msun) the angular velocity at the H-burning shell decreases by 25\%(from 8.75 to 6.62$\times 10^{-2}$\rps), the mass accreted increases by 13\% (from 1.27 to 1.46$\times 10^{-4}$\msun),  the ignition density decreases by 12\%, and the maximum value of $L_H$ decrease by 30\%. Moreover, at the epoch of the maximum extension of the flash-driven convective shell the H-rich zone is a factor 1.8 more massive than the accreted matter. These results demonstrate that the physical conditions of the H-ignition point (mainly density) and, hence, the strength of the resulting H-flash depend on the rotation rate of the accreted layers.
%------------------ Figure 6  -----------------------------------------------
\begin{figure}
	\centering
	\includegraphics[width=\columnwidth]{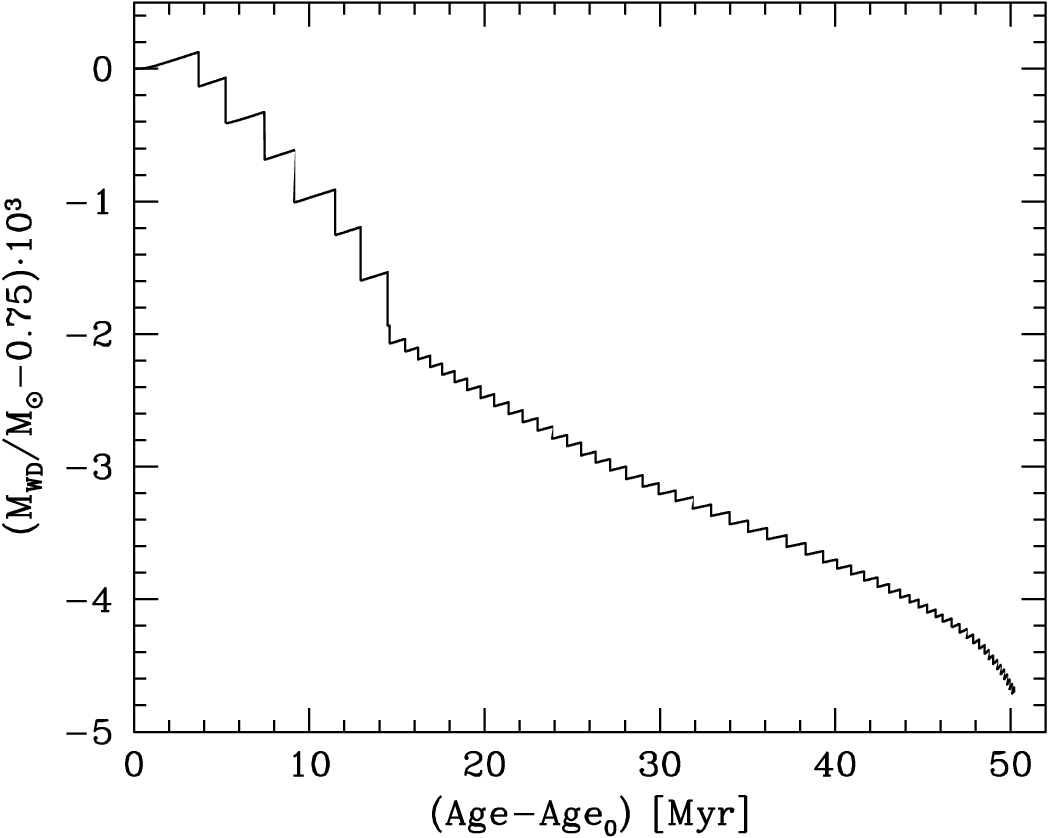}
	\caption{Time evolution of the WD total mass during the phase of H-rich mass accretion for the rotating model. Age$_0$ is the epoch of RLOF by the sdB star.
	}
	\label{fig6}
\end{figure}

As in the non-rotating model, due to the H-flash, the rotating WD expands and fills its  Roche lobe giving rise to mass loss from the binary system. As illustrated in Fig.~\ref{fig6}, where we report the evolution of the accreting WD total mass, all the matter accreted before the H-flash, as well as the most external layers of the original WD, are lost via Roche lobe overflow, resulting in the reduction of the WD mass. Figure~\ref{fig6} also discloses that the entire phase of H-rich matter accretion is characterized by a secular reduction of the initial WD mass. Nevertheless, the evolution of the mass transfer rate with time during the H-rich mass transfer phase  for the rotating model is very similar to that of the non-rotating WD, shown in Fig.~\ref{fig1}. In fact, even if in the rotating model the WD mass is reduced, its variation is small, so that the effect on the binary separation is practically negligible. 

A further inspection of Fig.~\ref{fig6} reveals that  the evolution of the WD total mass is very irregular. For instance the amount of mass lost by the WD is at maximum during the Roche lobe episode triggered by the seventh H-flash. Such an occurrence is due to the fact that during the previous six flash episodes the most external zones of the original WD have been removed, so that during the seventh episode the accreted matter starts to be mixed via rotation-induced instabilities with layers enriched in \isotope{12}{C} (X(\isotope{12}{C}) > 0.1), produced during the evolution of the CO WD progenitor in He-shell burning. This occurs when the H-flash is on the verge to be ignited, resembling very close the \textit{Mixing during the Thermonuclear Runaway} invoked by \citet{starrfield2020} to explain Nova outbursts \citep[see also ][]{casanova2011,jose2020}. As a consequence, the H-flash becomes stronger, the burning extends inward, engulfing a larger zone of the WD, which is subsequently involved in the mass loss via Roche lobe overflow.

In the following H-rich mass transfer episodes, the accreted matter is mixed with C-enriched material during the entire accretion phase, so that a smaller amount of mass has to be accreted to trigger the H-flash. This situation is quite similar to what occurs in non-rotating models accreting H-rich matter with increased CNO abundance. At the end, we note that for (Age-Age$_0$) > 43 Myr, when the hydrogen abundance in the accreted matter drops below 0.5 (see Fig.~\ref{fig1}), the mass transfer rate from the donor rapidly increases so that the compressional heating produced on the WD surface is larger and, hence, a smaller amount of mass has to be accreted to trigger H-flashes (see Fig.~\ref{fig1}).

As a whole, we find that due to the accretion of H-rich mass the CO WD experiences 62 flashes and its total mass reduces to \mwd=0.7453\,\msun.
%------------------ Figure 7  -----------------------------------------------
\begin{figure}
	\centering
	\includegraphics[width=\columnwidth]{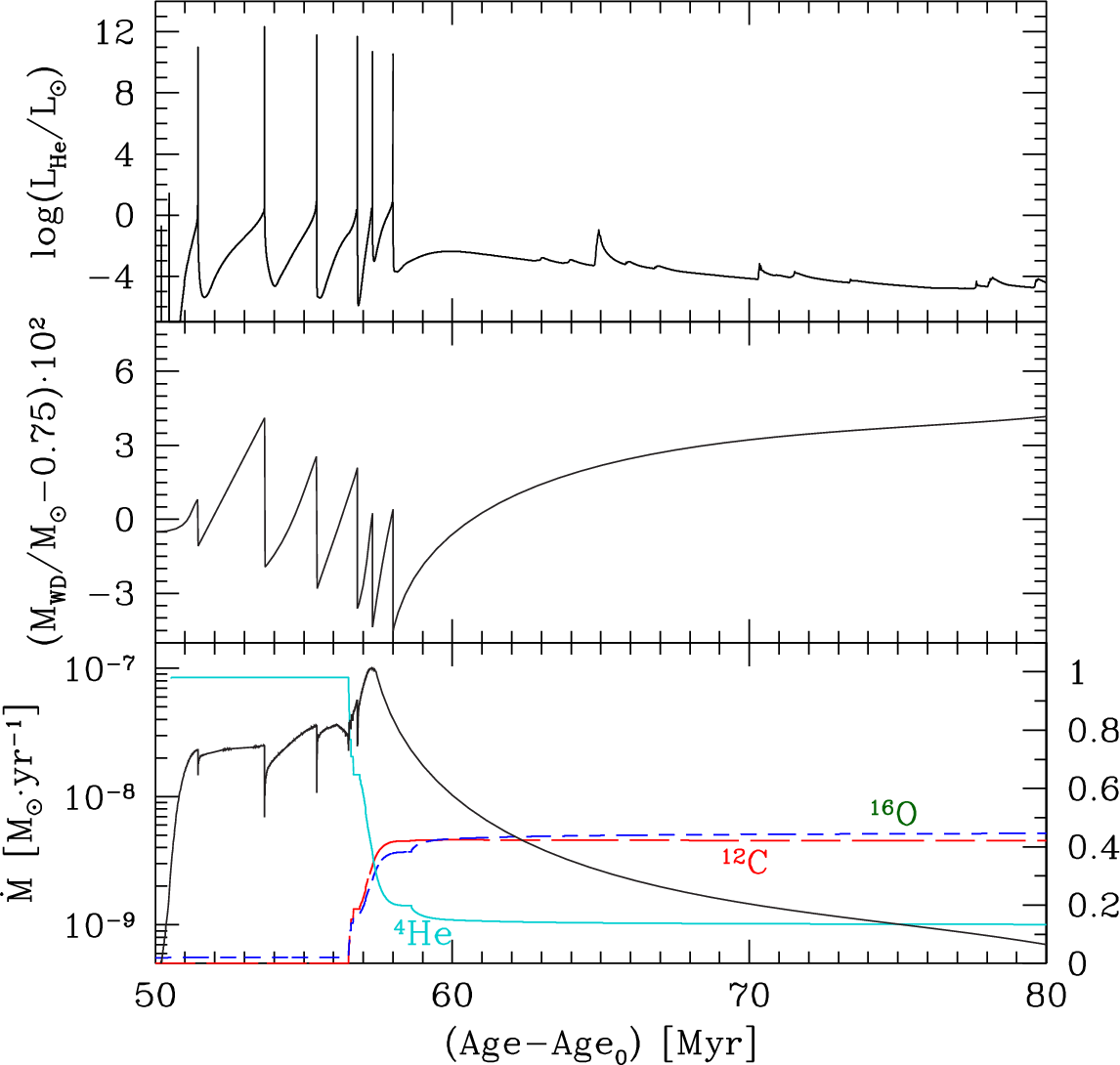}
	\caption{Time-dependence of the He-burning luminosity (top panel), WD total mass (middle panel) and mass transfer rate (lower panel, left y-scale) during the stage of He-rich matter accretion. The two leftmost spikes in the upper panel show, for comparison, the He-burning luminosity during the last two (61th and 62nd) H-burning outbursts.The lower panel also presents the evolution of the abundances of \isotope{4}{He} (solid cyan line), \isotope{12}{C} (long dashed red line) and \isotope{16}{O} (dashed blue line) as tracers of the chemical composition of the accreted matter (right y-scale). 
	}	
	\label{fig7}
\end{figure}

\subsection{Evolution during the He-rich matter accretion phase}
When the H-rich mantle of the sdB has been completely transferred, He-rich matter starts to be accreted and the corresponding \mdot rapidly increases above $\sim 10^{-8}$\,\msyrm\ (see lower panel in Fig.~\ref{fig7}). We find that our model experiences a strong, non-dynamical He-flash after $1.29\times 10^{-2}$\,\msun\ of He-rich matter has been accreted. However, as shown in Table~\ref{heflashes}, where we report several physical properties of the accreting WD during the He-flashes phase, due to rotation-induced mixing the mass of the He-rich zone, \textit{i.e.,} the one with He-abundance larger than 0.1 ($\Delta M_{He}^{pk}$ in Table~\ref{heflashes}), is almost two times larger than the mass of the accreted matter $\Delta M_{acc}$. 
As a consequence, during the Roche lobe overflow episode triggered by the He-flash all the accreted matter as well as the most external zone of the WD are lost from the binary system, 
so that the retention efficiency, defined as the percentage of transferred mass effectively retained by the accretor ($\eta=1-\frac{\Delta M_{lost}}{\Delta  M_{acc}}$), is negative. By computing the long term evolution of this model, we find that the CO WD in the \ptf\, binary experiences six He-flashes, as illustrated in Fig.~\ref{fig7}.
%
%------------------------------- Table 2 
\begin{table}
\caption{Selected properties of the accreting WD during the evolution throughout recurrent He-flashes of the rotating model. We report, for each flash episode, the mass of the WD at the beginning of the cycle (\mwd\, in \msun), the orbital period at the reonset of mass accretion (Per., in min.), the mass accreted between two successive flashes ($\Delta M_{acc}$ in $10^{-2}$\,\msun), the mass of the He-rich zone with X(\isotope{4}{He}) > 0.1 ($\Delta M_{He}^{pk}$ in $10^{-2}$\,\msun), the mass lost during the flash-driven Roche lobe overflow ($\Delta M_{lost}$ in $10^{-2}$\,\msun), the ignition density for He-burning ($\rho_{He}$ in $10^5$\gcc), the maximum temperature ($T_{He}^{max}$ in $10^{8}$K) and maximum luminosity ($L_{He}^{max}$ in $10^{10}$\,\lsun) attained during the He-flash, and the retention efficiency defined as $\eta=1-\frac{\Delta M_{lost}}{\Delta  M_{acc}}$ .}
\label{heflashes}      
\centering                   
\begin{tabular}{l c c c c c c }
\hline\hline                 
\# Fl.                &   1   &   2   &   3   &   4   &   5   &   6   \\ 
\hline
\mwd                  & 0.745 & 0.739 & 0.730 & 0.722 & 0.714 & 0.706 \\
Per.                  & 24.04 & 21.36 & 17.84 & 13.88 & 9.46  & 8.51  \\
$\Delta M_{acc}$      & 1.285 & 5.184 & 4.554 & 4.871 & 3.859 & 4.793 \\
$\Delta M_{He}^{pk}$  & 2.014 & 6.561 & 5.931 & 6.445 & 4.940 & 5.588 \\
$\Delta M_{lost}$     & 1.896 & 6.074 & 5.355 & 5.717 & 4.773 & 4.595 \\
$\rho_{He}$           &  1.24 &  2.51 &  1.88 &  1.90 &  1.56 &  1.97 \\                      
$T_{He}^{max}$        &  4.73 &  5.06 &  4.73 &  4.71 &  4.66 &  4.99 \\
$L_{He}^{max}$        &  1.21 & 87.05 & 37.49 & 45.24 &  4.35 &  0.17 \\
$\eta$                & -0.48 & -0.17 & -0.18 & -0.17 & -0.24 &  0.04\\
\hline                 
\end{tabular}
\end{table}
A further inspection of both Fig.~\ref{fig7} and Table~\ref{heflashes} reveals that the physical properties of the He-flashes do not exhibit a well defined regular behavior as a consequence of the variation in the mass accretion rate, chemical composition of both accreted matter and the matter in the external layers of WD, efficiency of angular momentum transport and temperature profile in the outermost zone of the WD. 

In particular, the amount of accreted mass triggering the first He-flash was lower than the one for the second outburst. The first flash is less energetic than the second one, despite 
\mdot in both cases is comparable immediately prior to the flash (see Fig.~\ref{fig7}). In fact, the He-abundance on the surface of the WD at the onset of the first He-rich mass transfer episode was 0.33, while at the onset of the second one it was only 0.14.  
During the last $\simeq 55$ H-flashes angular momentum remained stored in the most external layers of the WD, due to the  very short recurrence time of flashes which hampers its efficient inward transport. 
During the He-accretion phase, interpulse periods increase and angular momentum is transferred inward at a higher rate. For instance, at the onset of He-accretion the angular momentum was concentrated in the outermost $2.7\times 10^{-3}$\,\msun\ of the WD, while at the re-onset of accretion before the second flash such a zone was more than 10 times more massive ($3.9\times 10^{-2}$\,\msun).  This affects the strength of flashes, since angular momentum (and hence angular velocity) distribution defines the lifting effect of rotation, as well as the mixing efficiency. 
Finally, the temperature profile in the external layers of WD before He-accretion started was almost flat, with a very low peak of the order of 40 MK, while at the onset of the second mass-transfer episode the outermost $9\times 10^{-3}$\,\msun\ had a large thermal content with a peak temperature of almost 300 MK. 

To summarize, the structure of the outer layers of the WD at the beginning of the first and the second He-flashes are different and this produces differences in the properties of the flashes. From the second to the sixth He-flashes, the physical conditions in the outermost layers of the  WD at the re-onset of mass accretion after RLOF episodes are quite similar and the differences in the properties of the flashes depend on the mass transfer rate and/or the chemical composition of the accreted matter.

The third He-flash episode is characterized by a larger mass transfer rate as compared to the previous one, so that the mass accreted before the onset of the flash is lower and, hence, the resulting flash is also less strong. On the other hand, during the fourth He-rich matter accretion episode a larger amount of mass has been accreted as compared to the third one, even if \mdot increases to $5.7\times 10^{-8}$\,\msyrm, while the strength of the flash is comparable. In fact, during this accretion cycle, the zones of the donor partially involved in central He-burning start to be transferred onto WD, so that the He-abundance in the accreted matter decreases to 0.71\footnote{We remind that very soon after RLOF by low-mass He-stars, nuclear burning in their interiors quenches and the chemical composition  of transferred matter is defined by the extent of the sdB evolution prior to RLOF and the abundances profile  formed while nuclear burning extinguishes \citep{yung2008}.}.  As a consequence a larger mass has to be accreted to trigger the He-flash, since reduced X(\isotope{4}{He}) limits the rate of the 3$\alpha$-reaction and prevents a stronger flash.

During the fifth He-accretion cycle the mass transfer rate attains a maximum (\mdot$\simeq 10^{-7}$\,\msyrm). The He-abundance in the accreted matter decreases to 0.38, but unlike the previous flash episode, the increase in \mdot is now the leading parameter, resulting in a sizable reduction of the accreted mass.  At the end, during the sixth accretion episode, the mass transfer rate decreases to $4.5\times 10^{-8}$\,\msyrm\, resulting in an increase in the amount of matter to be accreted to trigger the He-flash, which is ignited at higher density, thus producing a maximum temperature in the He-rich buffer definitively larger. 
As the amount of available fuel above the He-burning shell is lower\footnote{During this mass transfer episode the helium abundance in the accreted matter decreases to 0.21.}, the energy injected into the He-rich envelope is lower and, as a consequence, during the following Roche lobe episode a small amount of the accreted matter is retained by the accreting WD. 
All in all we find that during the the six He-flashes the binary system loses about 0.2869\,\msun\, and the mass of the donor reduces to 0.1416\,\msun. 
%------------------ Figure 8  -----------------------------------------------
\begin{figure}
	\centering
	\includegraphics[width=\columnwidth]{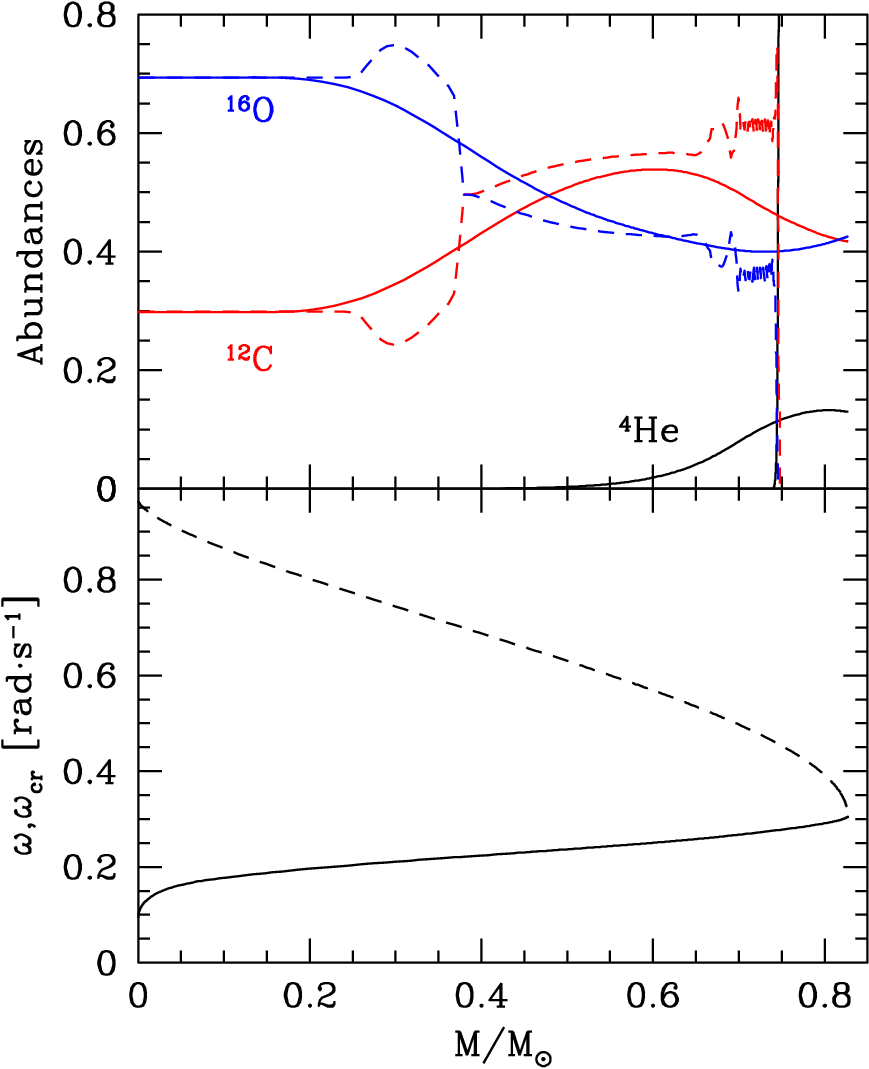}
	\caption{For the last computed structure in the rotating model, we report in the upper panel the abundance of \isotope{4}{He}, \isotope{12}{C} and \isotope{16}{O} (solid lines) and in the lower panel the $\omega$ and $\omega_{cr}$ profiles (solid and dashed lines, respectively). For comparison, in the upper panel we report the  abundances of the same isotopes in the initial 0.75\,\msun\, WD model (dashed lines).}	
	\label{fig8}
\end{figure}

After the sixth He-flash, when mass transfer from the donor resumes, the orbital period of the system is 10.16 min. The subsequent evolution of the system is similar to that of AM~CVn stars with strongly evolved non-rotating He-star donors at the beginning of RLOF \citep[see, \textit{e.g.},][]{yung2008,2010MNRAS.401.1347N}. It reaches the minimum of the periods at $\approx8.27$\,min and then evolves to longer ones. Mass-transfer rate continuously decreases and hence the amount of matter to be accreted to trigger a new flash becomes larger and larger and, as a matter of fact, it is never attained. We stop computations when the mass of the donor reduces to 0.018\,\msun\, because the adopted equation of state does not provide thermodynamic quantities for extremely cool objects. At that epoch, the orbital period is $\approx$34.47\,min. and the donor, corresponding to the homogeneous former core of a He-burning star, consists of a C/O mixture with traces of He. The accreting WD has now a mass \mwd= 0.8274\,\msun, while the mass transfer rate is \mdot=$2.17\times 10^{-11}$\,\msyrm. In Fig.~\ref{fig8} we report, for the last computed structure of the accretor, the profile of most abundant isotopes (upper panel) and that of $\omega$ and $\omega_{cr}$ (lower panel).

\subsection{Nucleosynthesis during He-flashes}

\begin{figure}
	\centering
	\includegraphics[width=\columnwidth]{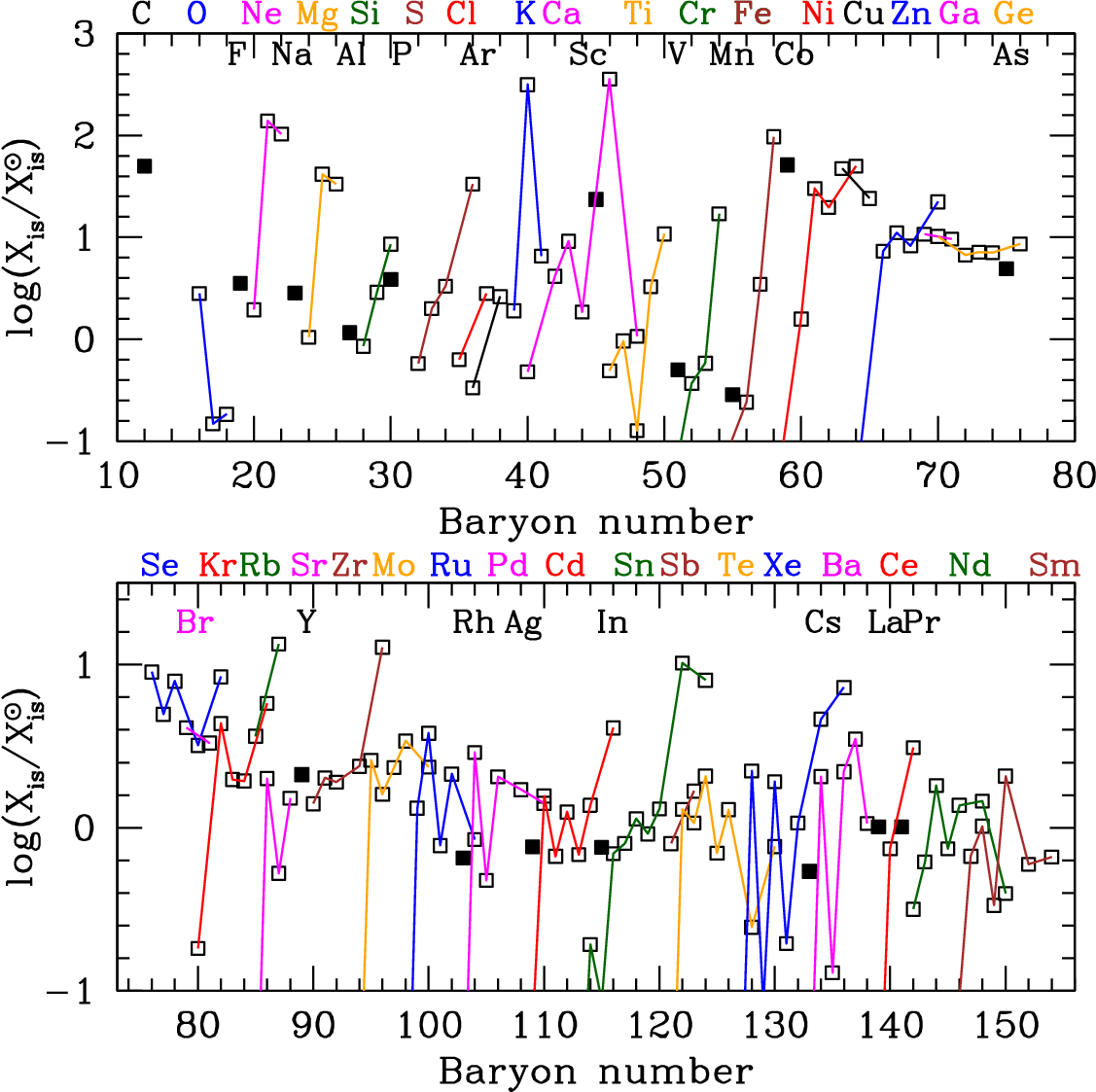}
	\caption{Abundances of isotopes up to samarium produced during the third He-flash episode in the rotating model relative to their solar value. The unstable isotopes have been allowed to decay to their stable isobars.}	
	\label{fig9}
\end{figure}

During the He-rich matter accretion phase, the resulting six flashes are very strong, so that the temperature in the He-burning shell exceeds $4\times 10^8$ K (see Table~\ref{heflashes}). Then, the \isotope{22}{Ne}$(\alpha,n)$\isotope{25}{Mg} reaction is fully efficient and, hence, $n$-capture nucleosynthesis is expected to occur. We recall that the \isotope{22}{Ne} abundance is practically coincident with the original CNO abundance in the progenitor of the sdB in \ptf. Isotope \isotope{22}{Ne} in the WD appears as a result of burning of nitrogen in the accreted matter via the reaction chain \isotope{14}{N}$(\alpha,\gamma)$\isotope{18}{F}$(\beta^+)$\isotope{18}{O}$(\alpha,\gamma)$\isotope{22}{Ne} during the initial phases of the He-flash. To characterize the chemical composition of the matter ejected by the system during the six RLOF episodes, we consider the third He-flash episode as representative of the whole phase and we repeat the computation by adopting a full nuclear network up to \isotope{210}{Po}, as in \citet{piersanti2019}. The results are displayed in Fig.~\ref{fig9}, where we report the abundance of isotopes produced during the He-flash episode relative to their solar value. As it can be seen, the most overproduced isotopes are \isotope{40}{K}, \isotope{46}{Ca}, \isotope{21}{Ne}, \isotope{22}{Ne} and \isotope{58}{Fe} whose abundances are larger than the solar ones by factors 314, 356, 138, 103 and 97, respectively. Isotopes with atomic numbers $26< Z\le 32$ are overproduced by factors larger than 10, while the abundances of isotopes from Br to Sm are increased by factors of a few. Finally, isotopes with Z> 62 do not exhibit any particular trend. On the other hand, we find that the abundances of \isotope{210}{Pb}, \isotope{210}{Bi} and \isotope{210}{Po}, the endpoints of the adopted nuclear network, are $1.2\times 10^{-11},\ 1.7\times 10^{-13},\ {\rm and\ }2.6\times 10^{-11}$, respectively, clearly indicating that the nucleosynthesis during the He-flash should produce also heavier isotopes and eventually also actinides. This topic is far beyond the aims of the present work and will be addressed in a forthcoming paper devoted to study the nucleosynthesis in interacing binaries. 
%------------------ Figure 9  -----------------------------------------------

\section{The final fate of \ptf\, system}\label{s:fate}
In our model of \ptf, when we stop the accretion of He-rich matter, the WD is stable because its angular velocity profile has attained an equilibrium configuration, so that the fate of this rather massive WD is to cool down. However, if angular momentum is removed from the star, then the latter can contract producing a compressional heating of the inner zones, thus potentially triggering the ignition of 3$\alpha$-reaction. \cite{chandra1970} and \cite{friedman1978} suggested that rapidly rotating compact stars can emit gravitational waves, thus becoming unstable for non-axysimmetric perturbations. Currently, two different modes are considered as responsible for these instabilities, namely the $f$-mode and the $r$-mode with nodal numbers of spherical harmonics $l=m=2$. In general, \cite{andersson2001} suggested that gravitational instabilities are strongly suppressed in the presence of large turbulence-driven viscosity, so that it is usually assumed that during the mass accetion no GWR emission could occur at all. In this regard, it has to be noticed that according to \cite{imamura1995} the secularly unstable modes could survive also in presence of shear instability. 

As shown in many works \citep[\textit{e.g.} see ][]{ostriker1969,durisen1975,durisen1981}, the $f$-mode instability in a rapidly rotating object can occur only if the ratio of the rotational energy and the gravitational one $\Theta=E_{rot}/|W|$ becomes larger than a critical value which is fixed to 0.14 for a wide range of angular velocity profiles, even if \cite{imamura1995} have shown that this limit could be as low as 0.09 for strongly differential rotating object. Our massive WD at the end of the mass transfer process has
$\Theta=0.044$, so we can exclude that this instability could affect at all the evolution of the WD angular momentum. 
On the other hand, $r$-modes are always unstable in rotating inviscid stars \citep{andersson1998} and their growth timescale can be estimated according to \cite{lindblom1999} as
\begin{equation}
	\tau_r=\left[{\frac{2\pi}{25}\left(\frac{4}{3}\right)^8 \frac{G}{c}} \int_0^R \rho(r)\left(\frac{r\omega(r)}{c}\right)^6 dr\right]^{-1},
\end{equation}
where $\rho(r)$ and $\omega(r)$ represent the density and angular velocity profile of the equilibrium configuration. 
Such a timescale has to be compared to the viscous timescale $\tau_v$ defined as
\begin{equation}
	\tau_v={\frac{{\frac{1}{5}\int_0^R\rho(r) r^6dr}}{\int_0^R\eta(r) r^4 dr}},
\end{equation}
where $\eta(r)$ represents the shear viscosity profile \citep{lindblom1999}. 
We find that, after the sixth He-flash, when the mass-transfer rate in our model of \ptf\, drops below $\sim 3.3\times 10^{-8}$\,\msyrm, \textit{i.e.} when the WD total mass grows above 0.71\,\msun, $\tau_v$ becomes larger than $10^{9}$ yr and rapidly increases above $10^{10}$ yr. 
For the final structure we computed, we obtain $\tau_r=2.36\times 10^9$ yr, an order of magnitude smaller than the viscous timescale. We remark that this value is, however, definitively longer than the one derived by \cite{yoon2004a} for a 1.5\,\msun WD (see their Fig. 14), mainly because in their structure the average angular velocity is a factor 5 larger than in our case. 

We compute the further evolution of the rotating WD by subtracting at each timestep $\delta t$ an amount of angular momentum equal to $\delta_j=J_{WD} \cdot \exp({-\delta t/\tau_r})$ where $J_{WD}$ is the actual value of the WD total angular momentum. We find that, as angular momentum is subtracted from the surface, a local gradient of the angular velocity forms, thus determining the re-onset of shear instabilities  in the outermost zones of the WD. As the braking process goes on, shear instability affects zones more and more internal, so that the whole structure progressively reduces its angular velocity; at the same time, also rotation-driven mixing resumes, and He-rich material is mixed downward at a higher level. In any case, as the braking timescale is very long, the WD does not heat up to any significant level and, hence, He-burning can not be ignited at all \citep[see also the discussion in ][]{yoon2004a}. 

With the aim of determining under which physical conditions He-burning could be eventually ignited in the accretor of \ptf, we compute exploratory models by adopting a braking timescale $\tau_b$ definitively shorter than $\tau_r$ above, even if no physical mechanisms responsible for such a fast slow-down has been identified/suggested so far. We find that for $\tau_b=2\times 10^7$ yr a very strong He-flash occurs at the mass coordinate 0.743\,\msun, the maximum attained temperature being of the order of $6\times 10^8$ K. For longer braking timescales the thermal energy delivered by the induced compression is transferred inward, without producing an increase in temperature in the He-rich layer large enough to trigger He-ignition. On the other hand, if $\tau_b$ is reduced, He-burning is ignited in progressively more external zone. As a matter of fact, He-detonation does not occur at all, as it was expected because  the He-buffer is not massive enough ($\Delta M_{He}\simeq 0.12$\,\msun) and, in addition, the He-abundance there is as low as 0.13 (see the lower panel in Fig.~\ref{fig8}).

\section{Discussion}\label{s:discuss}
 
Our results show that, due to the combined effect of rotation-induced lifting, shear dissipation heating and chemical mixing, the accreting WD in \ptf\, will never attain physical conditions suitable for developing a detonation and will end its life as a CO core capped by a massive He-buffer. Moreover, we find that during the mass transfer phase of both H-rich and He-rich material the WD experiences recurrent strong novalike flashes triggering Roche lobe overflow episodes during which all the accreted matter as well as a part of the underlying zones are lost from the system. As far as we know, such an occurrence has been never reported in the extant literature, even if \citet{yoon2004d} suggested that it could happen for low \mdot values. In fact, this is the first time that a series of H- and He-flashes have been computed for a rotating WD which accretes at low \mdote, so that during each runaway the accretor overfills its Roche lobe. The explanation of the secular reduction of the accretor
mass pulse-by-pulse should be sought in the mixing of the accreted layers with the underlying WD as determined by rotation-induced instabilities. Thus, the zone involved in the flash is larger as compared to the non-rotating model and, hence, the amount of mass that has to be lost during the Roche lobe overflow episode is larger.   
%
%------------------------------- Table 3 
\begin{table}
	\caption{Selected properties of models computed by varying the values of the parameters affecting the mixing efficiency of rotation-induced instabilities ($f_\mu,\ f_c$ - models T1 and T2, respectively), the amount of angular momentum deposited by the accreted matter ($f_j$ model T3) and by very high efficiency for the angular momentum transport (model T4). Computations have been performed for the third full cycle during the He-rich accretion phase ($\rm{ M_{WD}^{init}}$=0.7303\,\msun -- see Table~\ref{heflashes}). For the sake of comparison we report data also for the standard case. The accreted mass $\Delta M_{acc}$, the mass of the He-rich zone with X(\isotope{4}{He}) > 0.1 $\Delta M_{He}^{pk}$ and the mass lost during the flash-driven Roche lobe overflow $\Delta M_{lost}$ are in $10^{-2}$\,\msun, the WD final mass after the RLOF episode $\rm{ M_{WD}^{fin}}$ is in \msun, the ignition density for He-burning $\rho_{He}$ is in $10^5$\,\gcc, the maximum temperature attained during the He-flash $T_{He}^{max}$ is in $10^{8}$\,K, the maximum luminosity during the He-flash $L_{He}^{max}$ is in $10^{12}$\,\lsun, the retention efficiency $\eta=1-\frac{\Delta M_{lost}}{\Delta  M_{acc}}$.
	}           % title of Table
	\label{params}      % is used to refer this table in the text
	\centering                          % used for centering table
	\begin{tabular}{l c c c c c}        % centered columns (4 columns)
		\hline\hline                 % inserts double horizontal lines
		\# Model                &  STD   &   T1   &   T2   &   T3   & T4    \\ 
		\hline
		$f_\mu$                 &  0.05  &  0.005 &  0.05  &  0.05  & 0.05  \\
		$f_c$                   &  0.04  &  0.04  &  0.4   &  0.04  & 0.04  \\
		$f_j$                   &  1.0   &  1.0   &  1.0   &  0.6   & 1.0   \\
		$\Delta M_{acc}$        &  4.554 &  4.548 &  5.771 &  7.387 & 9.684 \\
		$\Delta M_{He}^{pk}$    &  5.931 &  5.956 & 11.513 &  8.866 & 10.060\\
	    $\Delta M_{lost}$       &  5.355 &  5.344 &  6.532 &  7.821 & ---   \\
	    $\rm{ M_{WD}^{fin}}$    &  0.7223&  0.7223& 0.7227 & 0.7260 & ---   \\
		$\rho_{He}$             &   1.88 &   1.88 &   2.76 &   3.46 & 4.36  \\
		$T_{He}^{max}$          &   4.73 &   4.73 &   5.21 &   5.43 & 5.72  \\
		$L_{He}^{max}$          &   0.37 &   0.38 &   0.80 &  11.20 & 77.29 \\
		$\eta$                  &  -0.18 &  -0.17 &  -0.13 &  -0.06 & ---  \\
		\hline                                   %inserts single line
	\end{tabular}
\end{table}

This is unavoidable as it is a direct consequence of rotation, even if, on  general grounds, the exact amount of matter lost during each flash episode depends on the assumed amount of angular momentum deposited onto the WD, as well as on the assumed efficiency of angular momentum transport and rotation-induced mixing. 
To investigate this issue we repeated the computation of the third He-accretion cycle by varying the scaling parameters: $f_j$ in Eq.~(\ref{eqjm}), which  determines the amount of angular momentum accreted onto the WD, $f_\mu$, defining  the efficiency of ``$\mu$-current'' opposing the rotation-induced circulation (see Section~\ref{s:numerics}), and $f_c$, affecting the efficiency of chemical mixing by rotation-induced instabilities (see Eq.~\ref{diffc}).  The results are reported in Table~\ref{params}.  

As it can be seen, a reduction of the $\mu$-barrier to rotation instabilities (model T1) produces only minor differences in the properties of the flashing structure. On the other hand, an increase in the mixing efficiency due to rotation-induced instabilities (model T2) causes a more rapid smearing of chemical gradient, so that angular momentum is transferred inward more efficiently, determining a reduction of the lifting effects due to rotation in the accreted He-rich buffer and, hence, a definitively larger inward extension of the He-rich zone. As a consequence, a larger amount of matter has to be accreted to trigger the flash, which also becomes stronger. Thus, the mass lost during the following Roche lobe episode is larger, even if the original WD is eroded practically at the same level.

At the end, the reduction of angular momentum deposited by the accreted matter (model T3) results in a lower rotation velocity in the accreted layers \citep[see for instance the discussion in ][]{yoon2004d}, so that a definitively more massive He-buffer is required to trigger the He-flash. Moreover, as the angular momentum gained by the accretor is lower as compared to the STD case, the effects of rotation are also more weak, so that the He-flash is definitively stronger even if the original WD is eroded at a lower level.

In the present work we did not account for the transport of angular momentum via magnetic instabilities. In general, a reliable physical theory describing the separate evolution of the poloidal and radial components of magnetic field is still missing. So, also in 1D the computation of the torque exerted in differentially rotating stars is particularly complex and largely uncertain. In a differentially rotating body, the onset of magnetic field is expected and its amplification could occur if the velocity field due to convection could generate a dynamo. However, \citet{spruit2002} demonstrated that, in differentially rotating stars, the magnetic field is amplified by an instability in the toroidal field, most probably the Tayler instability, which could occur also in radiative regions. Currently, the applicability of the so-called ``Tayler-Spruit'' mechanism as well as the derived efficiency of angular momentum transport has been questioned by some authors \citep[see, for instance, ][]{maeder2004,denissenkov2007}. Recently, \citet{fuller2019} suggested that the dissipation of unstable magnetic field perturbations determines the saturation of the Tayler instability, so that a smaller amount of energy is dissipated and the resulting magnetic field on average attains larger amplitudes. As a consequence, the corresponding angular momentum transport is definitively more efficient than that derived by \citet{spruit2002}.
	
This issue, as well as the effects of magnetic fields in the evolution of interacting binary systems experiencing recurrent mass transfer episodes will be addressed in details in a forthcoming paper. Here we limit our study to the computation of an additional test model (model T4 in Table~\ref{params}), in which we assume that angular momentum transport due to magnetic instabilities is efficient enough to enforce rigid rotation along the whole accreting WD. We find that the mass to 
be accreted to trigger the He-flash is more than doubled with respect to the standard case, in agreement with \citet{neunteufel2017} (see their Figure 7). In fact, due to the very efficient angular momentum transport, the amount of angular momentum stored in the accreted layers largely decreases and, in addition, the angular velocity gradient (the so-called ``shear factor'') rapidly goes to zero. As a result, viscous shear and GSF instability are suppressed, so that the mixing of the accreted matter with the original WD material is avoided (note that in model T4 the $\Delta M_{He}^{pk}$ is only 4\% larger than $\Delta M_{acc}$). In addition, the heating of the accreted layers via dissipation of rotational energy also does not occur. Finally, since angular momentum does not remain stored in the accreted layers, the $\omega$ value at the WD surface does not attain its critical value and, hence, a larger amount of angular momentum could be acquired by the accreting WD. As a consequence, the only effect of rotation is the reduction of the local gravity along the whole WD (lifting) and, hence, the amount of accreted matter needed to trigger a flash becomes larger as compared to the STD model. Due to the large increase of the mass of the accreted matter, He-ignition occurs at an higher density, so that the resulting He-flash is definitively stronger even if it does not turn into a detonation. We stop our computation  when the WD fills its Roche lobe. It has to be remarked that, according to the extant formulations, magnetic instabilities can operate only if the shear factor is larger than a critical value. This implies that a small gradient in the $\omega$ profile survives, thus triggering both the release of thermal energy via viscous dissipation and a partial chemical mixing. Both these effects result in reducing a bit the amount of mass to be accreted for the ignition of the He-flash which should be less energetic. 
These considerations seem to suggest that, after including also the effects of magnetic instabilities, the model of the accretor in \ptf\,  will not experience a detonation, even if the physical properties of the accreting WD, in particular the exact value of the retention efficiency during subsequent RLOF episodes, could be significantly different.

\section{Conclusions}\label{s:conclu}

We investigated the evolution of the binary system \ptf, harboring a CO WD with an sdB companion. This object \citep{Kupfer2022}, as well as similar ones \citep[\textit{e.g.,} CD-30 - ][]{deshmukh2024}, have been addressed as progenitors of double detonation events, leading to low-luminosity peculiar-nucleosynthesis type Ia Supernovae. At variance with previous studies, we considered the fact that, due to mass accretion from the companion, the WD acquires angular momentum and we followed the evolution of the rotation velocity in the accretor interior, as determined by convection and rotation-induced instabilities.

Due to the combined effect of rotation-induced lifting, shear dissipation heating and chemical mixing, the WD in \ptf\, will never develop the physical conditions for a detonation during the accretion stage and will become a CO core capped by a massive He-buffer. As well, detonation will not happen during subsequent evolution driven by angular momentum losses from the WD.

Such a conclusion applies also to all those binary systems with a He-rich component having a mass transfer history similar to that described for our case study. In particular, in CD-30 the rate of mass transfer from the donor is expected to resemble quite closely that of our case study \citep[compare our Fig.~\ref{fig1} with Figure 7 in ][]{deshmukh2024}, so that also in this case we suspect that the system will end its evolution without producing an explosion of Supernova proportion.
\begin{acknowledgements}
	We thanks the anonymous referee for useful comments and suggestions. We are grateful to S.-C. Yoon and N. Langer for helpful discussions. L.P. and L.R. Yungelson acknowledge partial financial support from the INAF Minigrant 2023 \textit{Self-consistent Modeling of Interacting Binary Systems}. L.P. acknowledges partial financial support from the Italian MUR project 2022RJLWHN: \textit{Understanding R-process {\rm \&} Kilonovae Aspects (URKA)}. E.B. acknowledges partial support from the Spanish grant PID2021-123110NB-100 funded by MICIU/AEI/10.13039/501100011033 and by FEDER, UE.
\end{acknowledgements}

   \bibliographystyle{aa} % style aa.bst
   \bibliography{piersanti} % your references

\end{document}